\documentclass{PoS}
\pdfoutput=1
\usepackage{graphicx}
\usepackage{subfig}
\usepackage{amsmath, amsthm, amssymb, bm}

\title{Poisson statistics in the high temperature QCD Dirac spectrum}
\ShortTitle{Poisson statistics in the high temperature QCD Dirac spectrum}
\author{Tam\'as G. Kov\'acs%
       \thanks{Supported by EU Grant (FP7/2007-2013)/ERC n$^o$208740. We also
       thank the Budapest-Wuppertal collaboration for making their code 
       for generating the lattice configurations available to us.
       }\\
       University of P\'ecs\\
       E-mail: \email{kgt@fizika.ttk.pte.hu}}
\author{\speaker{Ferenc Pittler}\\%
       University of P\'ecs\\
       E-mail: \email{pittlerferenc@gmail.com}}

\abstract{
We analyze the eigenvalue statistics of the staggered Dirac operator above
$T_{c}$ in QCD with 2+1 flavors of dynamical quarks. We use physical quark
masses in our simulations.  We compare the eigenvalue statistics from several
parts of the Dirac spectrum with the predictions of Random Matrix Theory for
this universality class and with Poisson statistics.  We show that at the low
end of the spectrum the eigenmodes are localized and obey Poisson statistics.
Above a boundary region the eigenmodes become delocalized and obey Random
Matrix statistics. Thus the QCD Dirac spectrum with physical dynamical quarks
also has the Poisson to Random Matrix transition previously seen in the quenched
SU(2) theory.
}

\FullConference{ The XXIX International Symposium on Lattice Field Theory - Lattice 2011\\
July 10-16, 2011\\
Squaw Valley, Lake Tahoe, California}

\begin{document}
\section{Introduction}

The Dirac operator represents the effects of the quark fields on the gauge field. 
The determinant of this operator appears in the QCD partition function. This operator
depends on the gauge field, we are looking for the statistics of its eigenvalues.
The matrix of the Dirac operator is very large, its linear dimension scales with the
lattice volume. The inverse of the Dirac operator occurs in physically measurable 
quantities. Therefore the low eigenvalues of the Dirac operator are very important.
\\
In the chirally broken phase (below $T_{c}$) the statistics of the Dirac spectrum
can be described by Random Matrix Theory \cite{Verbaarschot:2000dy}. However, we want to 
analyze the statistics of the Dirac eigenvalues above $T_{c}$ in the chirally symmetric phase. 
In this case we do not know any model for describing the statistics of the Dirac eigenvalues.
In the first step we regard the Dirac operator as a general fluctuating matrix with the same 
global symmetry as QCD. From this viewpoint we have two extreme 
possibilities:
\begin{itemize}
\item Random Matrix type spectrum
\\
In this case typical fluctuations in the matrix elements can freely mix eigenvectors, 
the eigenvectors are extended. If we increase the volume of the lattice, then the spatial extension 
of these eigenmodes will increase too which means that they are delocalized. They fill almost the entire 
lattice volume. The eigenvalues have Random Matrix statistics, they are not statistically independent.
\item Poisson type spectrum
\\
In this case fluctuations in the matrix elements cannot mix eigenvectors, the eigenvectors
are localized. The spatial extension of these eigenmodes will not depend on the volume of the lattice.
These eigenmodes are localized to a certain place in the lattice.
The eigenvalues follow Poisson statistics in this case, they are statistically independent.
\end{itemize}
Now the question is whether the eigenmodes of the Dirac operator are delocalized Random Matrix type,
or localized Poisson type.
\\
We first summarize the recent results that appeared in the literature.
Above $T_{c}$ in the chirally symmetric phase the spectral density of the Dirac operator around zero 
vanishes \cite{Banks:1979yr} and Random Matrix Theory has predictions at such soft edge \cite{Forrester}. 
Lattice simulations did not find agreement with these predictions. On the other hand, they indicate 
bulk Random Matrix statistics for the full Dirac spectrum \cite{Pullirsch:1998ke}. This would implicate
that the eigenmodes of the Dirac operator are delocalized. However, in Ref.\ \cite{GarciaGarcia:2006gr} 
the authors showed that around $T_{c}$ the eigenvalue statistics changes from Random Matrix type
to Poisson type spectrum. In Ref.\ \cite{Kovacs:2009zj} the author argued that the first few eigenvalues
of the Dirac operator are localized. 
\\
\section{Simulation details}
We examine the Dirac spectrum by computing the first few hundred eigenvalues of the Dirac operator for 
each gauge configuration. We divided this spectral range into many parts, and we examined the eigenvalue 
statistics separately in each part of the spectrum. We performed this analysis in SU(2) quenched theory in
\cite{Kovacs:2010wx}. In that case we saw a transition between localized and delocalized eigenmodes in the
spectrum. We want to know if this transition also appears in QCD. For this reason we must include 
dynamical fermions and we have to simulate SU(3) gauge theory. We examine whether the determinant of the 
Dirac operator modifies the statistics of the low Dirac eigenvalues or not. Our data are based on SU(3) 
gauge theory with (2+1) flavors of dynamical quarks at physical quark masses. We set the temperature
to about $2.6T_{c}$, well in the chirally symmetric phase. We use the action of Budapest-Wuppertal group 
\cite{Aoki:2006br}. This involves the staggered Dirac operator with two levels of stout smearing and a 
Symanzik improved gauge action. 
We have lattices with two different lattice spacings and for each lattice spacing we have gauge configurations on several
volumes. For the details see Table~\ref{tab:extension}.
\begin{table}
\caption{\label{tab:extension}The extension in time, and spatial direction of our lattices}
\begin{center}
\begin{tabular}{c | c c c}

$N_{t}$
&
\multicolumn{3}{c}{$N_{s}$} \\
\hline
$4$
&16 &24  & 32\\
$6$
& 24 & 36 
\\
\end{tabular}
\end{center}
\end{table}
\section{Results}
\begin{figure}
\begin{center}
\includegraphics[width=0.6\columnwidth,keepaspectratio]{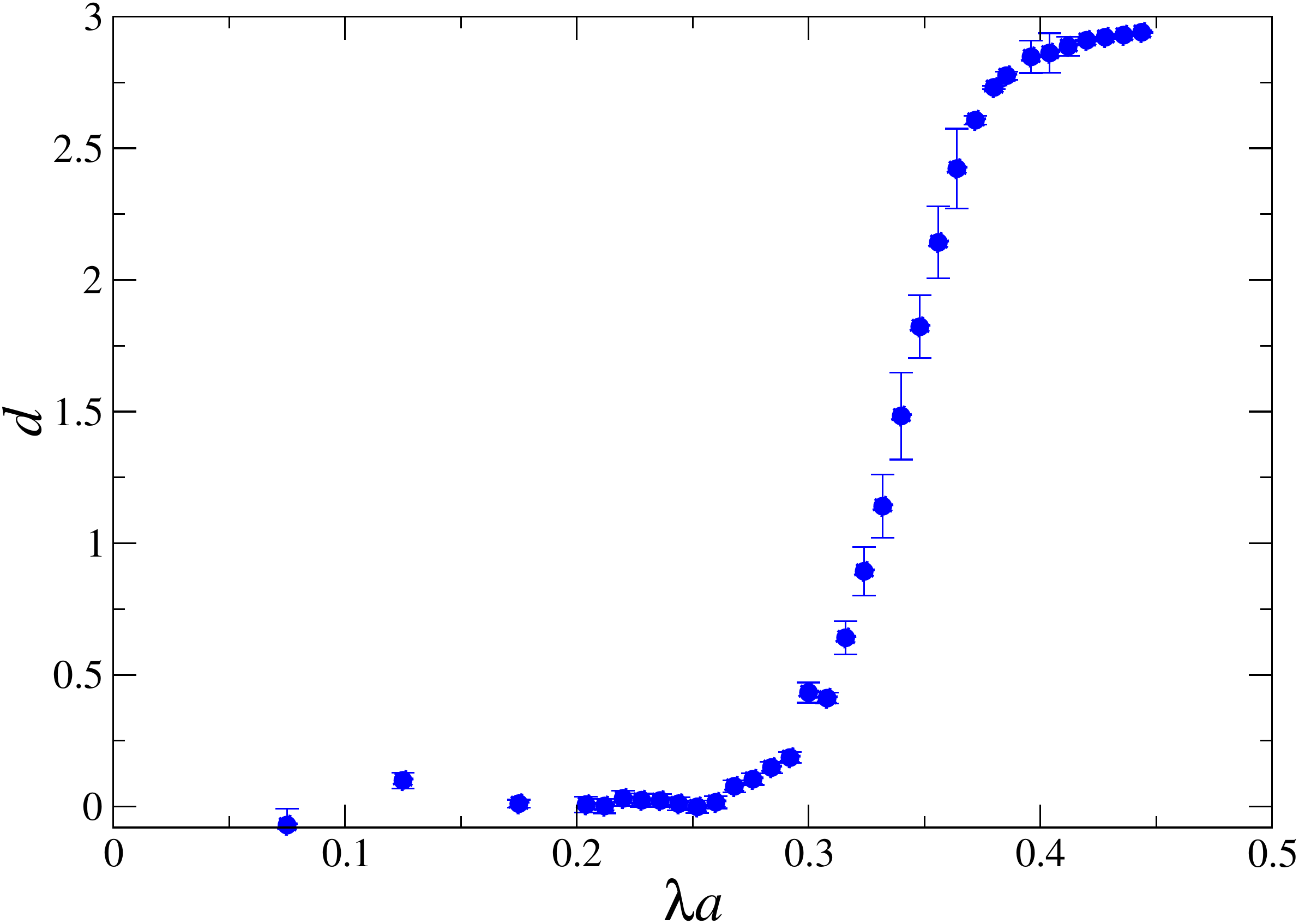}
\caption{\label{fig:em_dim}Effective dimension of the eigenvectors, $N_{t}=4$}
\end{center}
\end{figure}

\begin{figure}
\begin{center}
\includegraphics[width=0.7\columnwidth,keepaspectratio]{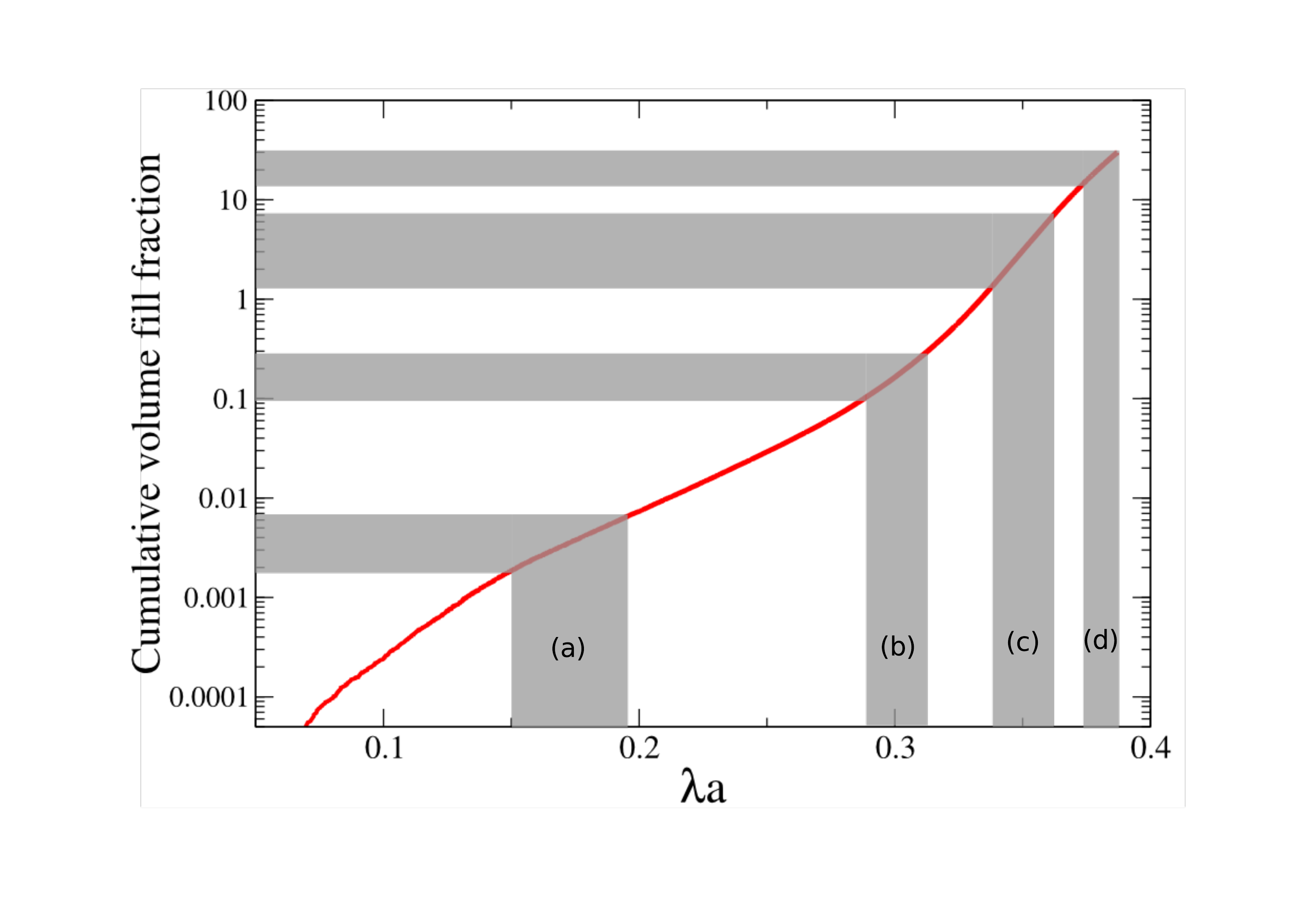}
\caption{\label{fig:cum_fill}Cumulative volume fill fraction $N_{t}=4,N_{s}=32$}
\end{center}
\end{figure}

Our main question is, whether the eigenmodes of the Dirac operator are localized, or delocalized. 
To answer this question we have to know how to measure the spatial extension of an eigenmode. 
We measure the volume of a general eigenmode with the help of this formula
\begin{equation}
{\cal V}=\left[\sum_{x:~lattice~site}\vert\psi_{i}\left(x\right)\vert^{4}\right]^{-1}
\label{eq:em_vol}
\end{equation}
This is equivalent to the product of the participation ratio and the total volume of the lattice.
To illustrate the usefulness of eq.\ \ref{eq:em_vol} we give a simple example. Let us have a normalized 
eigenmode $\psi_{i}\left(x\right)$ which spreads out uniformly in a volume ${\cal V}$. At a given site 
$x$ the absolute value squared of the wave function is equal to $\frac{1}{{\cal V}}$. 
A short calculation shows that in this special case the quantity defined in eq.\ \ref{eq:em_vol} yields
the volume occupied by $\psi_{i}(x)$.
Now the question is how this volume scales with the linear extension of the
lattice. Our first guess is that this is proportional to the $d$-th power of $L$, the lattice linear 
extension
\begin{equation}
{\cal V}\left(L\right)=C \cdot L^{d},
\end{equation} 
where $d$ is the effective dimension of the eigenmode. Now we will examine this quantity throughout 
the spectrum. In Fig.\ \ref{fig:em_dim} we display this effective dimension as a function of the 
eigenvalue. The eigenvalues are in lattice units and $d$ is the effective dimension of the average 
eigenvector corresponding to a certain eigenvalue range. At the low end of the spectrum this quantity 
is equal to zero which means that these eigenmodes are localized, their spatial extension does not 
change as we increase the lattice volume. Going up in the spectrum the effective dimension increases.
As we reach the bulk of the spectrum the dimension becomes roughly $3$ which means 
that the bulk eigenvectors spread out in all three spatial directions, they are completely delocalized. 
Between these two extremes there must be some transition.
\\
We examine the other quantity that determines how freely the eigenmodes can mix, the spectral density.
If the number of eigenmodes in a given eigenvalue range is small and these eigenmodes are localized,
they hardly mix. On the other hand if the number of eigenmodes is large in the same range, and the 
eigenmodes are delocalized, then they can freely mix. We want to measure both of these effects. Therefore,
we define a quantity that we call the cumulative volume fill fraction (CVFF). This is the sum of the volumes of all the eigenmodes up to a given eigenvalue 
$\lambda$, normalized by the total number of configurations. We measure this quantity in box size units.
We display the CVFF in Fig.\ \ref{fig:cum_fill}. 
At the low end of the spectrum this quantity is much smaller than unity. Therefore we expect, that these 
eigenmodes are independent, and they are produced independently in different subvolumes. Near 
$\lambda a=0.4$ this quantity becomes much bigger than unity. This means that these modes are strongly 
overlap.
\\
\begin{figure}
\begin{center}

\begin{tabular}{cc}
\includegraphics[width=0.4\columnwidth,keepaspectratio]{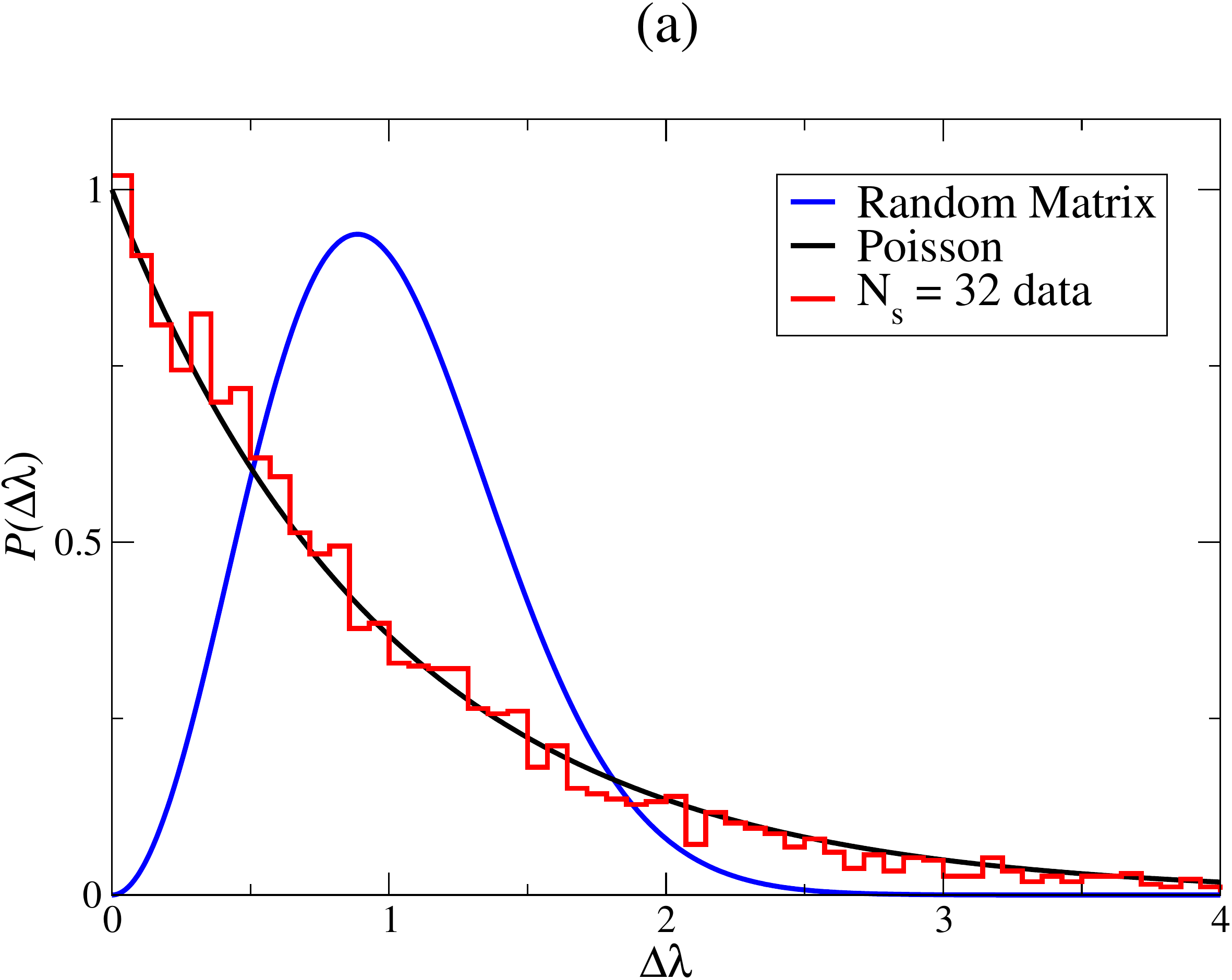}&

\includegraphics[width=0.4\columnwidth,keepaspectratio]{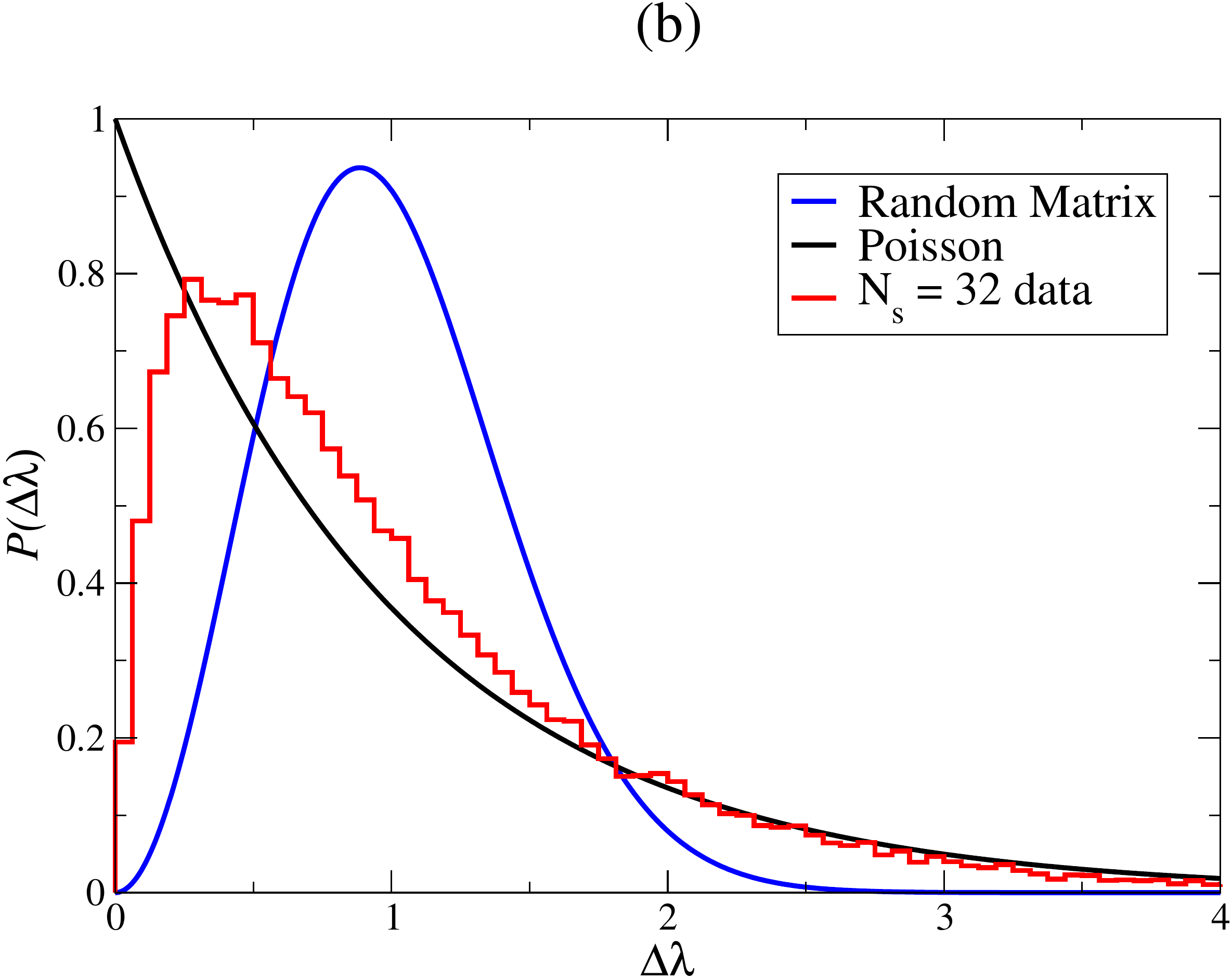}\\

\includegraphics[width=0.4\columnwidth,keepaspectratio]{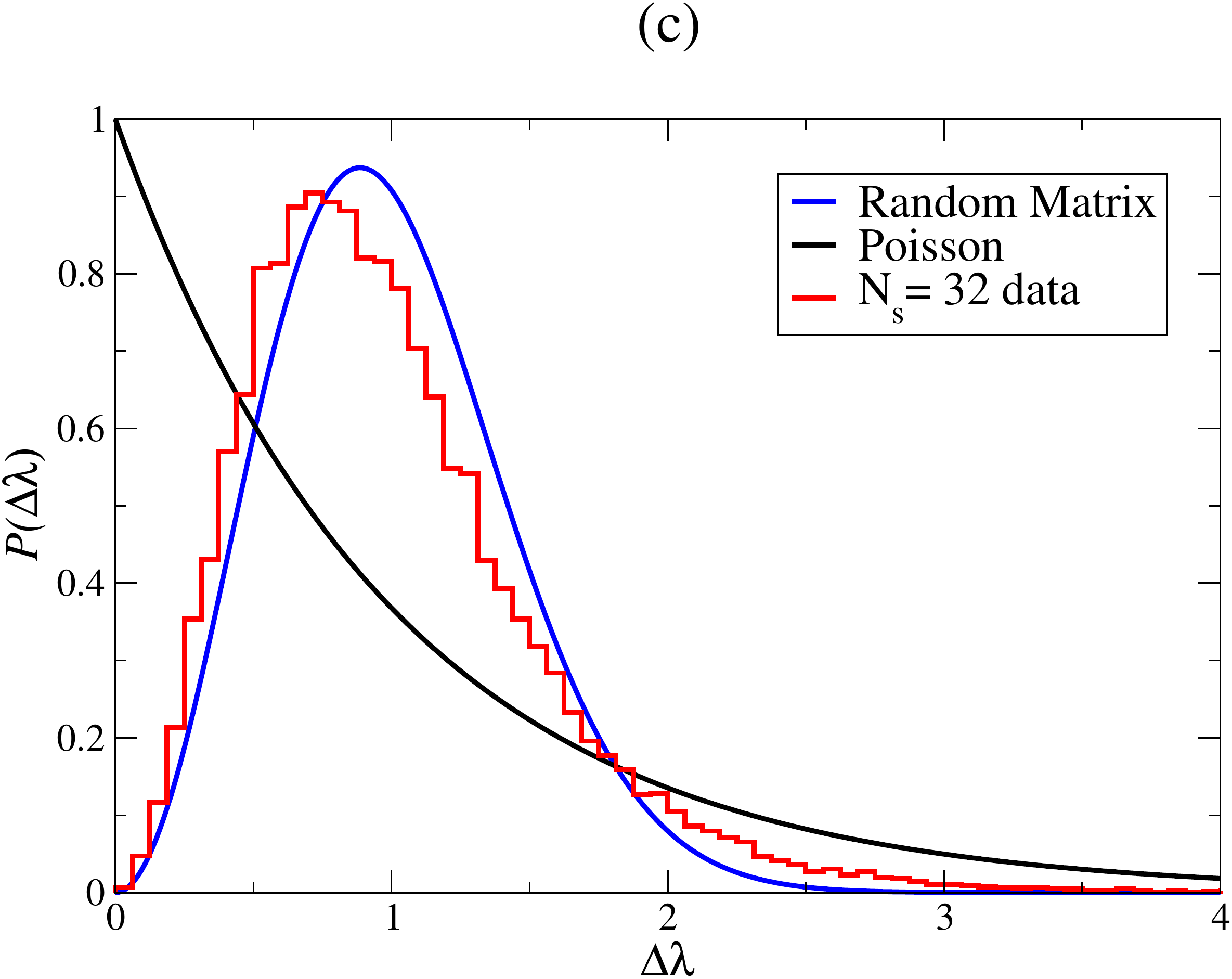}&

\includegraphics[width=0.4\columnwidth,keepaspectratio]{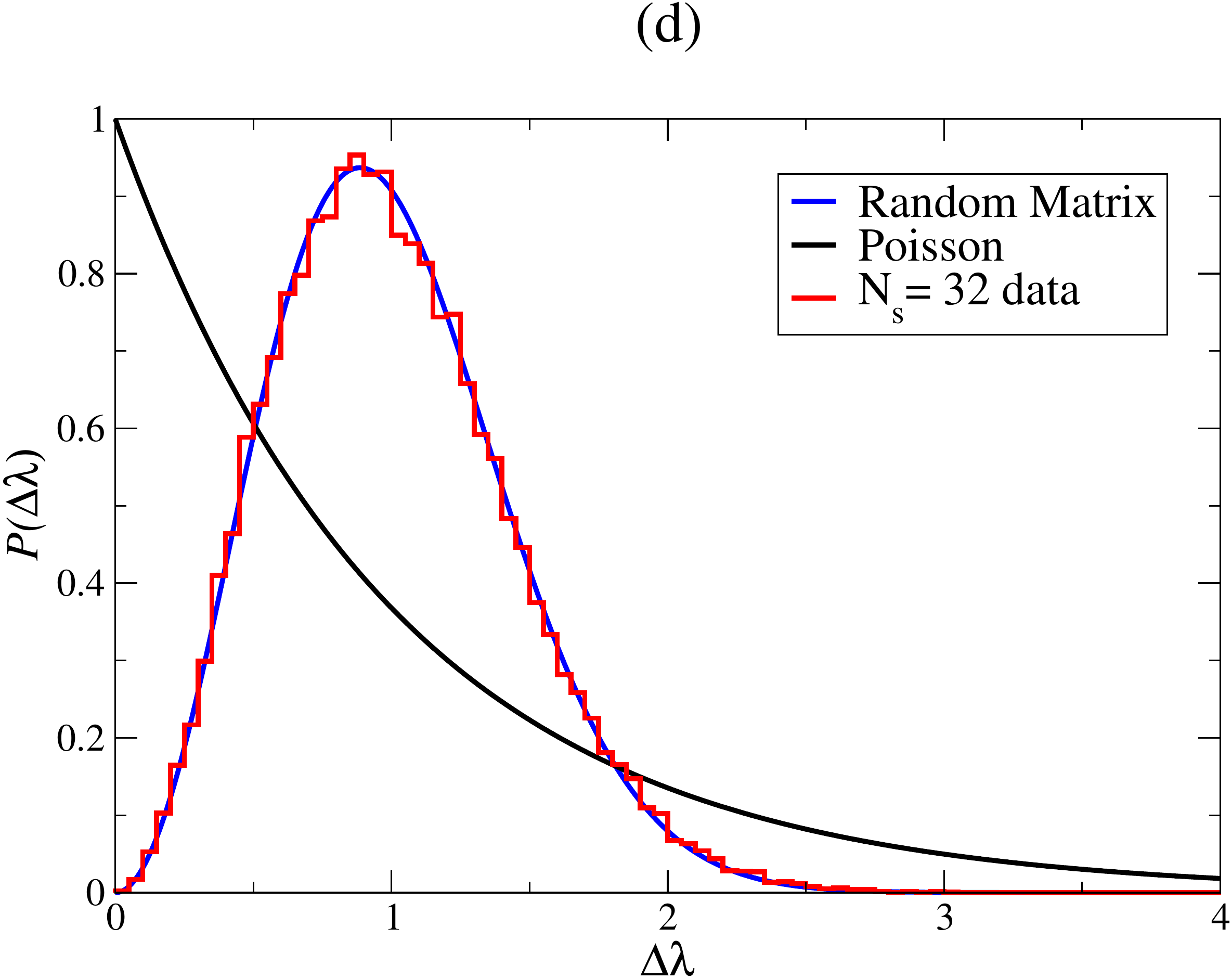}
\\
\end{tabular}
\end{center}

\caption{\label{fig:unfold}Unfolded level spacing distribution $N_{t}=4,N_{s}=32$,
$\Delta\lambda = \frac{\lambda_{n+1}-\lambda_{n}}
{\langle \lambda_{n+1}-\lambda_{n} \rangle}$}
\end{figure}
Next we examine the eigenvalue statistics separately in the four regions indicated in 
Fig.\ \ref{fig:cum_fill}. In order to compare the eigenvalue statistics with Poisson and Random Matrix 
type spectra we have to unfold the spectrum. After this transformation we retain only the universal 
correlations in the spectrum. To this end we rescale the eigenvalues to make the spectral density 
equal to unity throughout the whole spectrum. We have to choose some statistics which we can easily obtain from our data 
and for which we have analytical prediction in the Poisson and Random Matrix case. We computed the 
unfolded level spacing distribution (ULSD) of eigenvalues. For Poisson type localized eigenmodes the ULSD
follows a simple exponential. At the low end of the spectrum our result is in Fig.\ \ref{fig:unfold}(a).
We can see that our data match this exponential well, demonstarting that the eigenvalues are statistically
independent. 
Going up in the spectrum the statistics changes. (Fig.\ \ref{fig:unfold}(b),Fig.\ \ref{fig:unfold}(c)).
For Random Matrix type delocalized eigenmodes the ULSD can be described by the Wigner surmise of the 
corresponding random matrix ensemble. In the bulk of the spectrum in Fig.\ \ref{fig:unfold}(d) we can 
see that our data match perfectly this Wigner surmise. In this case the eigenvalues are not statistically 
independent, instead they repel each other. We see a transition from Poisson type localized eigenmodes to 
Random Matrix type delocalized eigenmodes in this case too.
\\
\begin{figure}
\begin{center}
\includegraphics[width=0.4\columnwidth,keepaspectratio]{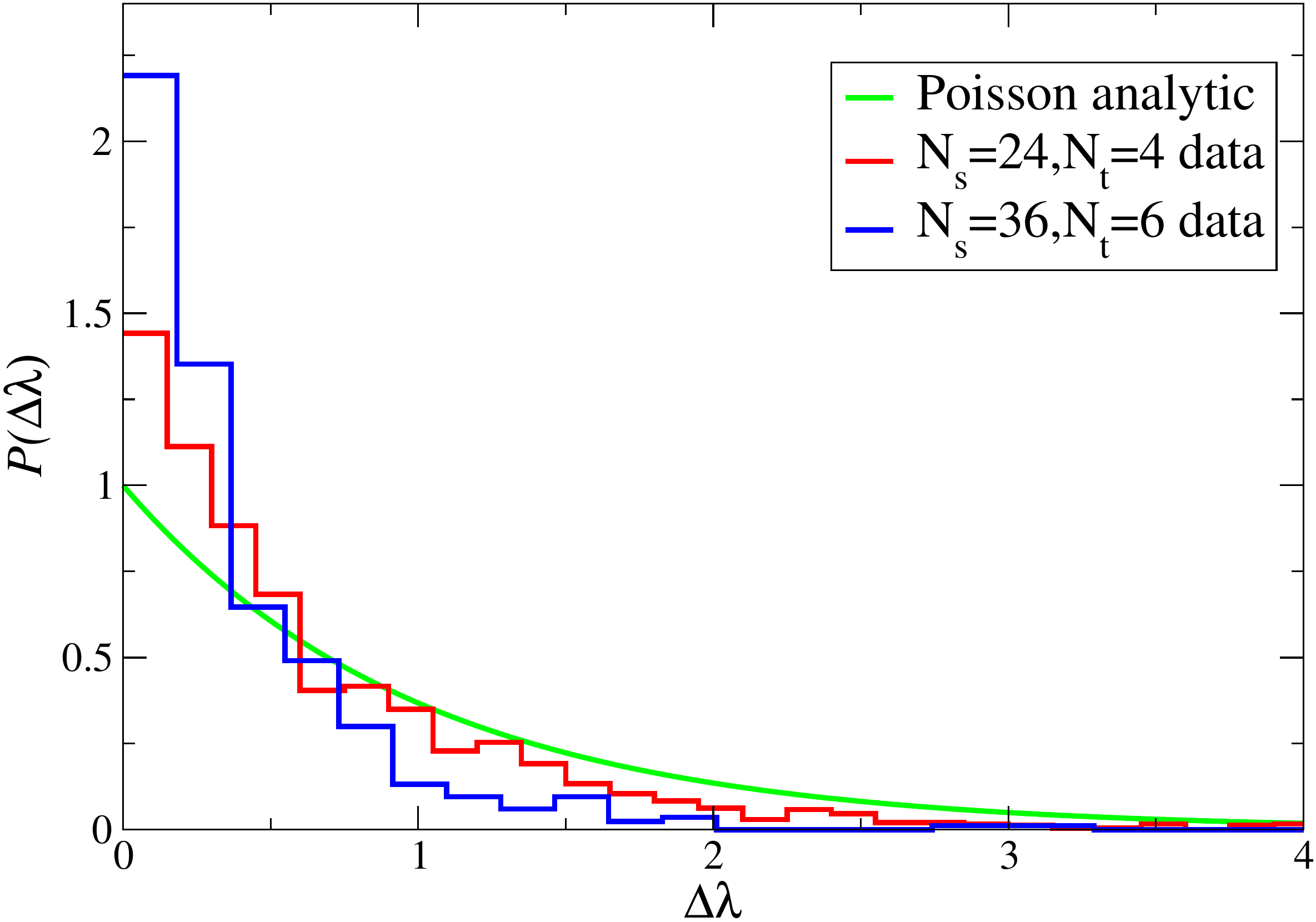}
\caption{\label{fig:stag_deg_a}ULSD between the first two eigenvalues } 
\end{center}
\end{figure}
\begin{figure}
\begin{center}
\includegraphics[width=0.4\columnwidth,keepaspectratio]{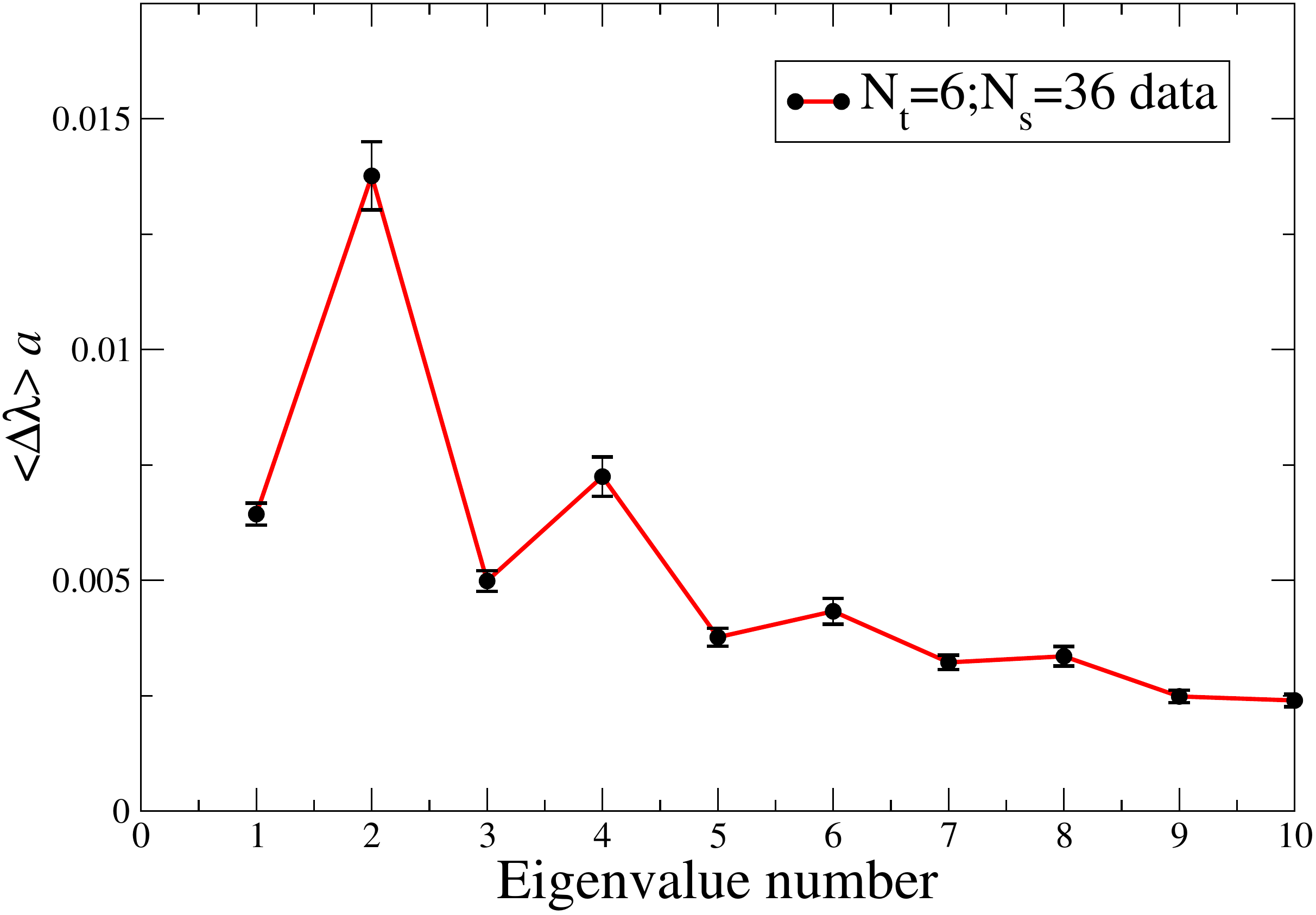}
\caption{\label{fig:form_d}Average level spacing of the first few eigenvalues} 
\end{center}
\end{figure}
Zooming on the lowest part of the spectrum reveals that the statistics of the lowest 
Dirac eigenvalues differs from the Poisson statistics. We examined the ULSD between the first two 
eigenvalues (Fig.\ \ref{fig:stag_deg_a}). We saw that the first two eigenvalues 
are not independent, instead they attract each other. Going to our finer lattices, we can see 
that this effect becomes stronger. We do not see such an effect on our simulation with SU(2) quenched
gauge theory with the overlap Dirac operator. We want to know what is the cause of this effect. We used 
staggered fermions, and we know that the eigenvalues of the staggered Dirac operator are four fold 
degenerate in the continuum limit. We used two levels of stout smearing which reduces the mixing between
high frequency modes of the theory. Therefore the spectrum of the Dirac operator looks more 
"continuum-like". In this case the staggered eigenvalues start to form a doublets. This can easily be 
seen in Fig.\ \ref{fig:form_d} where we plot the average level spacing corresponding to $n$-th eigenvalue. We see that there are two relevant scales in the problem:
\begin{itemize}
\item
splitting inside doublets
\item
spacing between doublets
\end{itemize}
\begin{figure}
\begin{center}
\includegraphics[width=0.4\columnwidth,keepaspectratio]{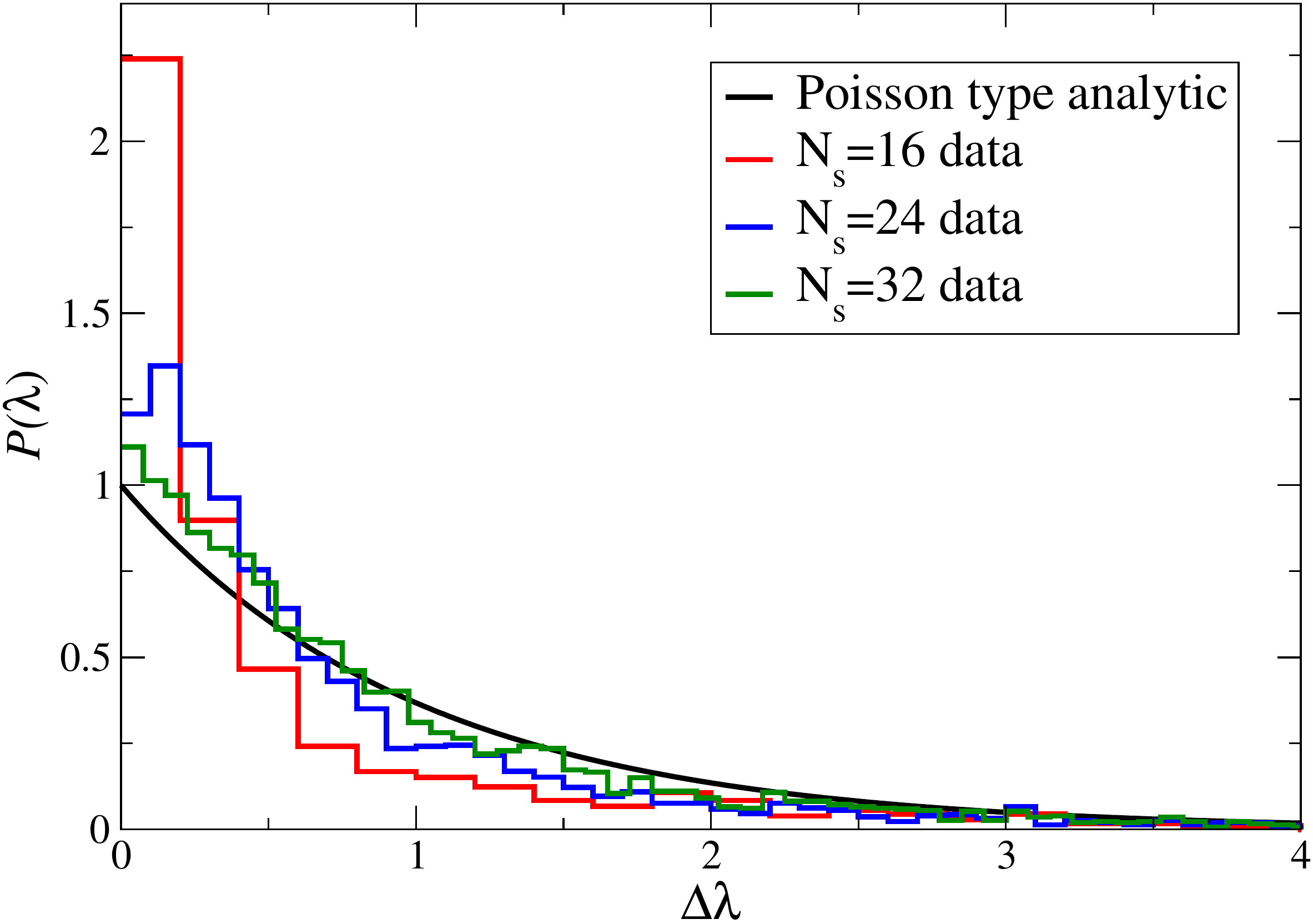}
\caption{\label{fig:last_plot}ULSD in the same range for different volumes} 
\end{center}
\end{figure}
The splitting inside doublets does not depend on the volume, but the spacing between doublets 
will decrease when we increase the volume. Therefore if the volume is large enough we cannot see 
this effect at a certain lattice spacing, because if the spacing between doublets becomes
comparable to the splitting, we cannot see doublets any more. This effect can be seen 
in Fig.\ \ref{fig:last_plot}. 
Examining the level spacing distribution at the very low end of the spectrum for three different
volumes shows that this effect becomes weaker in larger volumes.
\section{Conclusion}
We have seen that the fermionic determinant does not destroy the transition in the spectrum from 
Poisson type localized eigenmodes to Random Matrix type delocalized eigenmodes. We saw the transition
on our finer lattice too. We see that the transition remains also in the thermodynamic limit.
Because we use the staggered Dirac operator, the eigenvalue statistics on the low end of the spectrum is
disturbed by the doublets, but this effect is only a finite volume artefact.
\\

\end{document}